\documentclass{aastex}

\shorttitle{NIRSPEC spectra of Liller~1 and NGC~6553}
\shortauthors{Origlia et al.}

\begin{document}

\title{High resolution infrared spectra of bulge globular clusters: \\
Liller~1 and NGC~6553}

\author{Livia Origlia}
\affil{Osservatorio Astronomico di Bologna, Via Ranzani 1,
I--40127 Bologna, Italy}
\email{origlia@bo.astro.it}

\author{R. Michael Rich}
\affil{Math-Sciences 8979, 
Department of Physics and Astronomy, University of California
at Los Angeles, Los Angeles, CA 90095-1562}
\email{rmr@astro.ucla.edu}

\author{Sandra Castro}
\affil {Palomar Observatory 105-24, California Institute of Technology,
Pasadena, CA 91125}
\email{smc@astro.caltech.edu}

\altaffiltext{1}{Data presented herein were obtained
at the W.M.Keck Observatory, which is operated as a scientific partnership
among the California Institute of Technology, the University of California,
and the National Aeronautics and Space Administration.
The Observatory was made possible by the generous financial support of the
W.M. Keck Foundation.}

\begin{abstract}

Using the NIRSPEC spectrograph at Keck II, we have obtained 
echelle spectra covering the range $1.5-1.8~\mu \rm m$ for
2 of the brightest giants in Liller~1 and NGC~6553, old metal
rich globular clusters in the Galactic bulge.
We use spectrum synthesis for the abundance analysis, and
find $\rm [Fe/H]=-0.3\pm0.2$ and $\rm [O/H]=+0.3\pm 0.1$
(from the OH lines) for the giants
in both clusters.  We measure strong lines for the alpha elements
Mg, Ca, and Si, but the lower sensitivity of these lines to
abundance permits us to only state a general $\rm [\alpha/Fe]=+0.3\pm0.2$
dex.  The composition of the clusters is similar to that
of field
stars in the bulge and is consistent with a scenario in which
the clusters formed early, with rapid enrichment.
We have difficulty achieving a good fit to the spectrum of
NGC~6553 using either the low or
the high values recently reported in the
literature, unless unusually large, or no $\alpha$-element
enhancements are adopted, respectively.

\end{abstract}


\section{Introduction}

With the growing awareness that the bulge globular clusters
are a distinct subsystem has
come a new urgency to determine their ages and chemical composition.
To explore possible connections between clusters and the field, 
it is most interesting to compare the detailed
compositions of these clusters with
the Galactic bulge field
population (McWilliam \& Rich 1994).
As simple stellar populations, these clusters also will be play an 
important role in population synthesis for giant elliptical galaxies.  
At issue is determination of the their iron and alpha 
element abundances from integrated light.

For many of these bulge clusters, foreground extinction is so great
as to largely preclude optical studies of any kind, particularly 
at high spectral resolution.  
With the availability of NIRSPEC, a high throughput infrared echelle
spectrograph at the Keck Observatory (McLean et al. 1998), comes the prospect 
of measuring the composition of the clusters in the bulge.  

In this paper we present new, high resolution NIRSPEC spectra 
of bright giants in Liller~1 and NGC~6553. 
Both are relatively massive globular clusters $(M_V \sim -7.5)$
lying 2.5 kpc from the Galactic Center (Harris 1999).
This is the first in a series of papers which will address these 
globular clusters, 
the last major globular cluster subpopulation in the Galaxy for which high 
resolution abundances remain mostly to be determined.

Because of the relatively low extinction toward NGC~6553, this cluster
has been widely studied and it is one of the few bulge clusters  
with a very secure turnoff age measurement.
Ortolani et al. (1995) use WFPC2 
turnoff photometry to demonstrate that NGC~6553 and the similarly metal rich 
cluster NGC 6528 are coeval with the
well studied globular cluster 47 Tuc.  A proper motion
cleaned color-magnitude diagram of this cluster obtained using HST 
(Zoccali et al. 2001) cements the determination that this cluster is coeval 
with the Galaxy's oldest populations.
On the contrary, high extinction is present toward Liller~1.
Its turnoff is not very well defined 
(Ortolani et al.  2001) but it is also likely to be old.  
Accurate abundance measurements of these old clusters give crucial insight 
into the process of early chemical enrichment in the bulge of our Galaxy.

In the last few years NGC~6553 and Liller~1 have been the subject
of both optical and near infrared photometry (Ortolani, Bica \& Barbuy 1996, 
Guarnieri et al. 1998, Frogel, Kuchinski \& Tiede 1995, Davidge 2000).
Integrated spectra of their innermost core regions
in the optical (Bica \& Alloin 1986, Armandroff \& Zinn 1988) and
in the near infrared (Origlia et al. 1997; Frogel et al. 2001) are also
available, while only for NGC~6553 
have optical spectra been obtained for individual
red stars at both low (Minniti 1995; Coelho et al. 2001)
and high resolution (cf. e.g. Barbuy et al. 1992; 1999, Cohen et al. 1999, 
Carretta et al. 2001).

Despite the relative large number of photometric and spectroscopic
measurements, particularly in NGC~6553 (cf. e.g. Table 5 of Barbuy et al. 1999 for
a summary), the inferred abundances are still somewhat discrepant, leaving open 
the question of the cluster actual metallicity.
In particular the two most recent optical studies at high spectral resolution 
for members of NGC~6553 (cf. Barbuy et al. 1999, Carretta et al. 2001)
disagree by 0.5 dex. 

The near-IR spectral range is particularly suitable to study
cool, metal rich and possibly heavily reddened stars of the
bulge population.
The importance of the infrared CO and OH bands
as reliable carbon and oxygen abundance tracers in red giants
had been known for a few decades (cf. e.g Thompson et al. 1969;
Thompson \& Johnson 1974, Lambert et al. 1984, Kleinmann \& Hall 1986,
Wallace \& Hinkle, 1996).  
More recently, such a diagnostics has been also used to study
the metal content of the integrated red stellar population in globular
clusters and early type galaxies (Origlia et al. 1997, Origlia 2000, Frogel
et al. 2001) and to infer the oxygen abundance of some metal poor
stars (Balachandran \& Carney 1996, 
Mel\'endez, Barbuy \ Spite 2001).

Our observations and the infrared echelle spectra follow in Sect.~2. 
Sect.~3 discusses our abundance analysis, while in Sect.~4 the inferred 
metallicities and radial velocities are compared with previous results.  
Our concluding remarks are given in Sect.~5.

\section{Observations and Data Reduction}

We obtained
near infrared, high-resolution echelle spectra of two bright giants
in the cores  of the bulge globular
clusters Liller~1 and NGC~6553, on 26 July 2000.
We used the infrared spectrograph NIRSPEC (McLean et al. 1998) which is
equipped with an ALADDIN 1024$\times$1024 InSb array detector and
mounted at the Keck~II telescope.
The spectrograph employs a single echelle grating in quasi-littrow mode with
a 5$^{\circ }$ out-of-plane angle and a lower resolution grating with a
smaller blaze angle and zero out-of-plane angle as
cross dispersor.
The slit has a width of $0\farcs43$ (3 pixels) and a length of 12\arcsec\ 
and the nominal resolving power is R=25,000 (i.e. 12 km~s$^{-1}$).
We employed the standard NIRSPEC-5 setting, which uses the 
H-band filter and covers most of the 1.5-1.8 micron
band with only small inter-order gaps.

The night of 26 July 2000 was photometric, but the dome rotation
failed, permitting observations only of objects crossing through
the dome slit, which fortunately lay nearly due South. 
We exposed for
$3\times 10$ min in Liller~1, and 
$2\times 10$ min in NGC~6553.
 It must
be assumed that a portion of the exposures were compromised by
some occultation of the pupil by the dome.
Considering that little time was available, it was
fortunate that 2 bright giants in each cluster fell
on the slit (without requiring image rotation). The
$12\arcsec$ long slit was long enough to permit nodding
between exposures.
The H band images of the slit are given in Fig.~1; these
were obtained using the slit viewing camera 
(SCAM) of NIRSPEC, which
is equipped with a PICNIC 256$\times $256 HgCdTe array
detector and has a field of view of 46\arcsec$\times$46\arcsec\
and a scale of
$0\farcs183$$~pixel^{-1}$.

The raw frames were background subtracted and flat fielded.
Each order of the echellogram has been then extracted and straightened
according to its tilt angle. 
The normalized one-dimensional
spectra have been obtained summing over the 3 brightest  
rows and dividing by the continuum, which was determined  
applying a low-pass smoothing filter through each spectrum.
Imperfect subtraction of some of the brightest OH night sky
lines leaves some apparently high continuum points.  Prior
to fitting the final continuum, we overplotted the night sky
spectrum and verified that these regions were spurious.
The atmospheric absorption
features have been removed using a reference O-star spectrum,
but these are not significant, in any case.
The inferred signal to noise (S/N) ratio of the final spectra is
always $\ge$40 and the measured instrumental FWHM is
$\approx$0.8~\AA.

The wavelength calibration uses the OH sky lines
(Oliva \& Origlia 1992); we compute a quadratic
spline in the dispersion direction with an overall calibration accuracy of
$\approx$0.16~\AA\ or equivalently $\approx$3~km~s$^{-1}$.
The reduced echelle spectra of the observed stars in Liller~1 and
NGC~6553, covering the 1.51-1.75 $\mu$m wavelength range, are shown in
Figs.~2 and 3, respectively. 
The spectra of the two stars in each cluster are practically identical within 
the noise.

\section{Abundance analysis}

We compute suitable synthetic spectra
of giant stars by varying the stellar parameters and the
element abundances, using an updated
version of the code described in Origlia et al. (1993) 
to obtain spectra in the whole 1-2.5 $\mu$m range. 
The computations  use the LTE approximation and are based 
on the molecular blanketed model atmospheres of
Johnson et al. (1980)
which have been extensively used for abundance
studies based on high resolution spectra of cool stars (e.g. Lambert et al. 1984;
Smith \& Lambert 1985, 1990).   
The atomic oscillator strengths and the excitation potentials have been taken 
from Kurucz's database. For the Ca lines around 1.615 $\mu m$ we used the 
values tabulated by Bi\`emont \& Grevesse (1973, hereafter BG73) 
which provide a more reasonable synthesis of the observed spectra 
(cf. Sec.~3.1 for a more quantitative analysis). 
The molecular oscillator strengths and excitation potentials have been 
computed as described in Origlia et al. (1993). 
The reference Solar abundances have been updated 
according to Grevesse \& Sauval (1999).

By using the CO ($\Delta v=3$) and OH ($\Delta v=2$) molecular lines in
the 1.5-1.8 $\mu$m spectral window we can infer
reliable carbon and oxygen abundances (cf. e.g. Lambert et al. 1984).
These lines are not
heavily saturated even in the high metallicity domain and still lie
on the linear part of the element curve of growth, where the
line equivalent width variation with the abundance is maximum.
We can also infer an estimate of the nitrogen abundance using the
CN ($\Delta v=-1)$ molecular lines (cf. e.g. Sneden \& Lambert 1982).

At the NIRSPEC resolution of R=25,000 we can measure several single 
rotaion-vibration OH lines. However,
our carbon abundances are mostly 
derived from CO bandheads; in contrast to OH, the CO lines are
more blended at our resolution. 

Other metal abundances can be derived from the atomic lines
of Fe~I, Mg~I, Si~I and Ca~I, altough heavily saturated and
somewhat less sensitive to the element abundance variation
compared to the molecular lines,
being on the damping part of the curve of growth, where
the equivalent width increases much more
slowly with the element
abundance.

The modeling of the main molecular and atomic lines for
cool stars in the H band is relatively straightforward.
The major source of continuum opacity is H$^-$ and it 
has its minimum near 1.6 $\mu$m. These two facts 
also minimize any dependence of the results on the choice of model 
atmosphere. 
The molecular lines are
due to rotation-vibration transitions in the ground
electronic state, so they can be safely treated under the
LTE approximation.  At these low temperatures, the metals
are mostly neutral or singly ionized, eliminating the need
to compute equlibria
for higher stages of ionization.  The atomic oscillator
strengths remain the major source of uncertainty, and their
computed values could require a detailed cross-calibration with
suitable stellar template spectra.
In Sect.~3.1 we discuss in some more detail how diferrent 
assumptions for the oscillator strength values can affect the 
line profiles and the derived element abundances. 

From the near infrared (IR) photometry of Liller~1 and NGC~6553
published by Frogel, Kuchinski \& Tiede (1995) and Guarnieri et al.
(1998), their E(J--K) reddening of 1.70 and 0.41 
and the corresponding A$_K$=0.87 and 0.24 corrections, 
and their distance moduli of $\mu _0$=14.7 and 13.6, respectively, 
we compute the (J--K)$_0$ colors and the absolute K magnitudes 
for the observed stars (cf. Table~1).
We use the color-temperature transformation and the bolometric corrections 
of Montegriffo et al. (1995)
specifically calibrated on globular cluster giants,
to derive the stellar temperature and bolometric magnitudes of the observed stars 
from the (J--K)$_0$ color (cf. Table~1).
The two giants in Liller~1 are the brightest in the cluster and
we infer T$_{\rm eff}\approx $3700 and 3900~K,
while those in NGC~6553 have T$_{\rm eff}\approx $3900 and 4000~K.
The expected uncertainty in the temperature estimate is $\le$200~K,
taking into account the possible systematic effects due to the reddening
correction and the color-temperature transformation.
Concerning the stellar gravity, according to the theoretical evolutionary tracks, 
in the high metallicity domain the expected  
values for the stars in the upper part of red giant branch are 
log~g$\le$1.0, depending on their actual luminosity and temperature
(see Origlia et al. 1997 and references therein for a more detailed 
discussion).

Both the photometric quantitites reported in Table~1 and the similarity of 
the infrared spectra (cf. Figs.~2 and 3) indicate that the two observed stars 
in each clusters should have very similar photospheric parameters.
Hence, as a first step in our analysis of the spectra,
we adopt (cf. Table~2) 
average temperature and gravity of
T$_{\rm eff}$=3800~K and log~g=0.5 for the
2 giants of Liller~1 and
T$_{\rm eff}$=4000~K and log~g=1.0 for those in NGC~6553.
We also adopt the microturbulence of $\xi=2~km~s ^{-1}$ 
according to the values inferred by Origlia et al. (1997)
from the CO bandheads in the integrated spectra of these
clusters.
In Sec. 3.1, we will consider how our derived abundances
depend on these parameters.

For both clusters
the best fits to the observed spectra using the above reference
stellar parameters have been obtained for
half Solar [Fe/H], Solar [O/Fe] and a similar
enhancement by a factor of $\approx $2 for the other
$\alpha $ elements (Si,Mg,Ca) (cf. Table~2).
We also find some $^{12}$C depletion and $^{14}$N enhancement
(by a factor of $\approx$2, as well), as expected from the 
first dredge-up mixing process
in the stellar interior during the evolution on the red giant branch
(cf. e.g. Boothroyd \& Sackmann 1999 for a recent review).
The fit also 
gives the [$^{12}$C/$^{13}$C]$\le$5 ratio. 
Such a low value cannot be explained by 
first dredge-up alone and requires some 
extra-mixing processes occurring during the last ascent
of the red giant branch
(Boothroyd \& Sackmann 1999 and references therein). 

An overall, conservative estimate of the uncertainty in the derived
absolute abundances is $\le$0.2~dex.
Our synthetic best fits superimposed on the
observed spectra of Liller~1 and NGC~6553
in three major spectral regions of interest,
are plotted in Figs.~4 and 5, respectively.
These synthtetic spectra well account for most of the observed features,
and the few major discrepancies should be mainly ascribed to a bad 
OH sky line and/or atmospheric feature subtraction. 

In such a high metallicity domain, it is almost impossible to find lines 
which are completely unblended, even at high resolution.
Nevertheless, we identified a few representative
metal and OH lines which are reasonably clean to provide 
useful equivalent width measurements. 
By integrating over a $\pm$1~\AA\ range from the line central wavelength 
we obtain values systematically larger (30$\pm$10\%) 
than those from a simple gaussian fit of the line using the instrumental  
0.8~\AA\ FWHM, possibly indicating that some blend effects are at work.
For this reason we decided to use the values from the gaussian fit
with an overall accuracy of $\pm$30~m\AA. 
These values are reported in Table~1.

The inferred heliocentric radial velocities (cf. Table 1)
for the stars in Liller~1 (66 and 61~km~s$^{-1}$)
are consistent with the value listed by Harris (1996) (52$\pm$15~km~s$^{-1}$).
In the case of NGC~6553 our velocities ($-9$ and $-5~km~s^{-1}$) are
well within the relative wide range of values (between -60 and +60 km~s$^{-1}$) 
published in the literature (cf. e.g. Table 3 of Coelho et al. 2001 for
a summary).

\subsection{Error budget}

In order to investigate the influence of slightly different assumptions
for the stellar parameters on the inferred abundances we also compute
synthetic spectra with $\Delta $A=$\pm$0.2~dex
(where A is the input element abundance for the computation of the 
line opacities and the variation refers to all elements),
$\Delta $T$_{\rm eff}$$=\pm$200~K, $\Delta $log~g=$\pm$0.5~dex and
$\Delta \xi$$=\pm$0.5~km~s$^{-1}$ with respect to the adopted reference
parameters (cf. Table~2).
The adopted ranges for the stellar parameters are
quite conservative and somewhat representative of
their maximum uncertainty.  
We analyze both the line profiles and the equivalent widths.
The resulting stellar spectra of the two clusters, 
centered on a few selected atomic and
molecular lines of interest are plotted in Figs.~6 and 7,  
while their measured equivalent widths are compared in Figs.~8 and 9, 
respectively .

A variation of $\pm$0.2~dex in the element
abundance affects only slightly the depth and the broadening
of the atomic line profiles, but strongly affects the 
more metal sensitive CO and OH lines.
The change in the line strength
due to a $\pm$0.2~dex abundance variation can be equally obtained by varying 
some other atmospheric parameters, but by quite a large amount.
For example, $\ge 0.5~km~s^{-1}$ variation in the microturbulent velocity, and also 
temperature variation of $\ge$200~K in the case of the Ca and Si lines, 
could have a similar effect, but larger (smaller) microturbolent velocities and/or 
lower (higher) temperatures are required to fit lower (higher) input abundances,
somewhat contrary to the stellar evolution prescriptions for population II stars
(cf. e.g. Straniero \& Chieffi 1991, Schaller et al. 1992, Bertelli et al. 1994,
Cassisi et al. 1999). 

The molecular bands behave in a similar fashion:
the major change in the line
strength due to a $\pm$0.2~dex abundance variation,
could also be the result of relatively large ($\pm$0.5~dex)
log~g variation in the case of the CO band heads, while for the
OH lines, somewhat more saturated, are mainly major temperature
and microturbulent velocity variations which affect the line profiles.
Again, as in the case of the atomic lines, a decrease in the input abundance 
has to be compensated for by lowering the stellar gravity and temperature or 
increasing the microturbulence velocity ({\rm viceversa} for an increase of 
the input abundance). These trends are in the opposite direction of 
the {\rm standard} evolution prescriptions along the red giant branch.  

Line equivalent widths are only marginal affected 
(within the $\pm$30~m\AA\ scatter) by the variation of the stellar 
parameters in the ranges mentioned above, the major scatter ($\le 
2\sigma$) with the element abundance.
In the case of the OH bands and the Si line (slightly blended with OH as well), 
a 200K increase in effective temperature with respect to the 
adopted reference gives model equivalent widths which 
are ($\ge 3\sigma$) lower than the observed widths.
Although one might be concerned
that the photometric estimate of the effective temperature can be quite uncertain 
(mainly due to reddening correction),
the molecular lines are expecially powerful in constraining its value.
The scatter between the abundances inferred from different molecular lines 
(cf. e.g. the case of the two selected OH lines) is even smaller than 
the adopted $\pm 0.2$ dex error, indicating that in presence of high signal 
to noise spectra the carbon and oxygen abundances
can be constrained down to a limit of $\pm 0.1$ dex.

It is also noteworthy that the 
stellar features under consideration show a similar trend
with variations in the stellar parameters, altough with different
sensitivities.
As a consequence of such a behavior, {\it relative } abundances are less
dependent on stellar parameter variations.
Moreover, the simultaneous fit of the different atomic and molecular bands 
in particular, requires more unique solutions for the adopted stellar parameters,
significantly reducing their initial range of variation and
ensuring a good self-consistency of the overall spectral
synthesis procedure.

On the basis of this analysis,
we can conclude that minor
stellar parameter variations in the proximity of
their best fit solutions
do not significantly affect (within $\approx$0.1 dex)
the inferred abundances, and certainly do not affect
the relative abundances.

As mentioned in Sect.~3, the adopted oscillator strength is an
additional source of uncertainty 
in the abundances derived from the analysis of the atomic lines.
In this respect, for some representative bright atomic lines in the H band 
we compared the results obtained using the oscillator strengths of 
the Kurucz's list with those empirically calibrated 
by Mel\'endez \& Barbuy (1999, hereafter MB99). 
On average, MB99 values are systematically lower but 
when the difference does not exceed 
0.2~dex (which is the case for most of the atomic lines 
under consideration), 
both the line profiles and equivalent widths are
only marginally affected and the overall scatter 
in the derived abundances is $<0.1$~dex.

As an example, the atomic parameters and the inferred equivalent 
widths of the Ca, Fe, Si and Mg lines shown in Figs. 6, 7 and 8 
(see also Table~1), adopting the Kurucz/BG73 or the MB99
oscillator strengths, respectively, for the two best fit models 
(cf. Table~2) to the observed spectra of Liller~1 and NGC~6553
are listed in Table~3 for comparison. 
The difference between the measured 
equivalent widths is generally
$\le 1\sigma (\le 30$~m\AA) with a small impact 
on the inferred element abundance ($\le$0.1~dex using the
Ca~$\lambda $1.61508, Fe~$\lambda $1.61532 and Si~$\lambda $1.58884 lines, 
and $\le$0.2~dex using the Fe~$\lambda $1.55317 and Mg~$\lambda $1.57658 
lines).
For the Ca$\lambda 1.61508$ line, 
if the larger oscillator strength 
by Kurucz (log--gf = 0.362) is used, we must assume
a significantly lower ($\le $0.5 dex) Ca  
abundance and an unlikely [Ca/Fe]$<$--0.2~dex underabundance.

\section{Discussion}

The red giant branch slope derived from
optical (Ortolani, Bica \& Barbuy 1996)
and near--IR (Frogel, Kuchinski \& Tiede 1995) color magnitude diagrams of
Liller~1 suggest a supra-solar metallicity, altough these estimates
rely on extrapolation of empirical relations calibrated on less metal rich
clusters.
Integrated spectroscopy based on the calcium triplet
(Armandroff \& Zinn 1988)
also suggests twice the Solar metallicity.
More recently, color-magnitude diagrams from near-infrared
photometry
(Davidge 2000), and integrated spectroscopy of the CO feature
around 1.6 $\mu$m (Origlia et al. 1997) are
both consistent with half solar metallicity.
Our present near IR high resolution spectra of the two single giants
confirm such a sub-solar value.
Davidge (2000) also finds that stars in Liller~1 are not as metal rich
as those in the sorrounding bulge.
Frogel, Kuchinski \& Tiede (1995) also noticed that
the stars in the sorrounding field of Liller~1 are redder than
the most likely cluster members but since they were not
able to solve the ambiguity between possible abundance or
redenning variations,
they did not use this external field to decontaminate the cluster
color-magnitude diagram.
Ortolani, Bica \& Barbuy (1996) do not perform a detailed
field decontamination of their optical color-magnitude diagrams,
however they suggest that the cluster stars can be as metal
rich as
the bulge population given the locus and curvature of the red giant branch.

The most recent photometric and spectroscopic estimates of the
metallicity of NGC~6553 range between one fifth (Cohelo et al. 2001) 
and almost Solar values (Carretta et al. 2001).
The two most recent sets of
high resolution spectroscopic measurements in the optical range
(Barbuy et al. 1999 and Cohen et al. 1999) give [Fe/H]=--0.55
by measuring 2 bright giants and [Fe/H]=--0.16
by measuring 5 red horizontal branch stars, respectively.
Caretta et al. (2001) propose a revised iron abundance
of [Fe/H]=--0.06 for NGC~6553, using the same spectra and
measured equivalent widths.
Our iron abundance estimate based on bright giants as well, is somewhat in
between the values proposed by Barbuy et al. (1999)
and by Cohen et al. (1999) as revised by Carretta et al. (2001).
An overall excess of $\alpha$-elements (by a factor of $\approx$2)
is found by all the high resolution studies.

Figures 10 shows that we cannot achieve a good fit using the
iron abundances recently reported in the literature.
We compare our observations with synthetic spectra, produced
for the two different iron abundances reported by Barbuy et al. (1999)
and by Cohen et al. 1999 (adopting
the suggested revision of Carretta et al. (2001)
and $\rm [\alpha/Fe]=+0.3$ in both cases).
Iron abundances as low as proposed by Barbuy et al. (1999)
or higher as proposed by Carretta et al. (2001) can still
be barely consistent with our observed iron line strengths,
but they would not optimize our fit.  
Moreover, as shown in Fig.~11 also the difference between the line equivalent 
widths as measured in the model and observed spectra
can exceed the 1$\sigma$ scatter.
Most significantly, both
measurements would be {\it inconsistent} with an
[$\alpha $/Fe] enhancement by +0.3~dex (found by both
groups, and present in the Galactic bulge field).
If [Fe/H]=--0.6, then we need to assume  $\rm [C/Fe]\approx 0.0$ 
and an [$\alpha $/Fe] enhancement $\ge $0.5~dex  to account for line 
strengths and equivalent widths
of the CO, OH and the other alpha elements,
while if [Fe/H]=--0.1,  a higher carbon depletion and only a 
marginal (if any) [$\alpha $/Fe] enhancement is required.  
Note that our finding of Solar oxygen abundance from the OH lines should
be indeed rather firmly settled within $\pm$0.1 dex (cf. Sect.~3.1).

The discrepancy in the inferred abundances among these different
data sets seem most likely due to systematic abundance 
analysis effects
rather than being related to the evolutionary stage of the observed stars
(cf. Coelho et al. 2001).
In this respect, it is 
also interesting to compare our oxygen abundance in NGC~6553 with that
of Cohen et al. (1999), where the 9 ev lines of the triplet 
at 7774\AA\ are used (with non-LTE corrections).
Cohen et al. find [O/Fe]=+0.50$\pm 0.2$, but each of their
four stars exceeds our
value of [O/Fe], with one star at [O/Fe]=$+0.68$.  
Nevertheless, these permitted oxygen lines of high-excitation potential seem to 
produce some artificial over-enhancement (by a factor of few) 
in the derived abundance compared to that one derived from 
the forbidden [OI] 6300\AA\ lines (cf. e.g.  Mel\'endez, \& Barbuy 2001
and references therein).
Considering
the importance of oxygen in chemical evolution, and
as an indicator of the star formation history, it will
be valuable to extend these comparisons over the full range of iron
abundance.

Our findings are consistent with recent abundance determinations for
field stars in the Galactic bulge (Rich \& McWilliam 2000), for
which $\rm [O/Fe]\approx +0.3$ at [Fe/H]=0 using 
the forbidden oxygen line
at 6300 \AA.  
According to the long-held paradigm that massive star supernovae mainly
produce alpha elements (cf. reviews by Wheeler, Sneden, \& Truran 1989 and
McWilliam 1997), our abundance patterns in NGC~6553 and Liller~1 add one more
piece of evidence consistent with early and rapid bulge formation (cf. e.g. Matteucci, Romano \& 
Molaro 1999, Wyse 2000).  

However, our findings for NGC~6553 and Liller~1 may not be
generally applicable to the bulge globular clusters; there is
clearly much work to be done.  
Consider the case of NGC~6528, which Ortolani et al. (1995) find to
have an age identical to that of NGC~6553.  The abundance
analysis of four red horizontal branch stars in NGC~6528
by Carretta et al. 2001 finds [Fe/H]=+0.07, with Solar oxygen
abundance and mild $\alpha-$ element enhancement.  The oxygen
abundance in NGC~6528 is measured from the 6300\AA\ forbidden
line and the carbon abundance is poorly constrained.  Nonetheless,
the need for more studies of bulge globular clusters is clear,
and the use of both optical and infrared spectroscopy could
help to settle some of the discrepant abundance determinations
(e.g. combining CNO abundances from the infrared with optical
measurements of other elements).

\section{Conclusions}

Using the NIRSPEC spectrograph at Keck~II, we have obtained H-band
echelle spectra at $R=25,000$ of two of the brightest giants 
in each of the Galactic
bulge globular
clusters NGC~6553 and Liller~1.  Employing spectrum synthesis, we
find [Fe/H]=$-0.3\pm0.2$ and [O/H]=$+0.3\pm 0.1$ for the giants
in both clusters. 
We measure strong lines for the $\alpha-$ elements
Mg, Ca, and Si, but the lower sensitivity of these lines to
abundance permits us to only state a general $\rm [\alpha/Fe]=+0.3\pm0.2$
dex.  The oxygen abundance is based on many lines in the OH spectrum,
rather than on the single forbidden line at 6300 \AA,
or the 9 ev triplet at 7774\AA.  Ours is the first use of the OH lines to
measure [O/H] in old, metal rich stellar populations.
Our finding that oxygen
is enhanced relative to iron is consistent with massive
star supernovae being responsible for the enrichment, and
it is noteworthy that alpha enhancements are securely
determined in old stars with a high iron abundance of [Fe/H]=$-0.3$.

Our abundances for these two ancient globular
clusters strengthens further the link
between the bulge globular clusters and the bulge field
population, pointing toward early formation and rapid enrichment
for both clusters and field.  

\acknowledgments

RMR acknowledges support from grant number AST-0098739,
from the National Science Foundation.
The authors are grateful to telescope operator 
Terry Stickel, whose competence and encouragment in the face
of technical difficulties made these observations
possible.  We are also grateful to the staff
at the Keck observatory and to Ian McLean
and the NIRSPEC team.
The authors wish to extend special thanks to those
of Hawaiian ancestry on whose sacred mountain
we are privileged to be guests.  Without their
generous hospitality, none of the observations
presented herein would have been possible.

\clearpage


\begin{deluxetable}{lccccc}
\footnotesize
\tablewidth{15truecm}
\tablecaption{(J--K)$_0$ colors, heliocentric radial velocity and
line equivalent widths (m\AA)
for the observed stars in Liller~1 and NGC~6553.}
\tablehead{
\colhead{}&
\colhead{Liller~1 \#1}&
\colhead{Liller~1 \#2}&
\colhead{}&
\colhead{NGC~6553 \#1}&
\colhead{NGC~6553 \#2}
}
\startdata
ref \#$^a$ & 1 & 2 & & 40201 & 47192 \\
(J--K)$_0^a$&  1.04 & 0.93 & & 0.93 & 0.86\\
M(K)$_0^a$& -6.91 & -6.28& & -4.87& -5.09\\
M$_{bol}^a$& -4.11 &-3.68 & &-2.27 &-2.59\\
$v_r$ [km~s$^{-1}$]& 66 & 61 & & --9 & --5\\
Ca~$\lambda $1.61508& 316 &310&  &274 &254\\
Fe~$\lambda $1.61532& 218 &243&  &237 &229\\
Fe~$\lambda $1.55317& 209 &186&  &196 &184\\
Mg~$\lambda $1.57658& 417 &418&  &435 &437\\
Si~$\lambda $1.58884& 505 &487&  &496&494\\
OH~$\lambda $1.55688& 307 &295&  &269 &251\\
OH~$\lambda $1.55721& 321 &307&  &288 &279\\
\enddata
\tablecomments{
$^a$ Stars in Liller~1 from Frogel, Kuchinski \& Tiede (1995),
in NGC~6553 from Guarnieri et al. (1998).}
\end{deluxetable}

\begin{deluxetable}{lccccc}
\footnotesize
\tablewidth{15truecm}
\tablecaption{Adopted stellar atmosphere parameters.}
\tablehead{
\colhead{}&
\colhead{Liller~1 \#1}&
\colhead{Liller~1 \#2}&
\colhead{}&
\colhead{NGC~6553 \#1}&
\colhead{NGC~6553 \#2}
}
\startdata
T$_{\rm eff}$ & 3800 & 3800 & & 4000 & 4000\\
log~g & 0.5 & 0.5 & & 1.0 & 1.0\\
$\xi$ [km~s$^{-1}$] & 2 & 2 & & 2 & 2\\
$\rm [Fe/H]$ & --0.3 & --0.3 & & --0.3 & --0.3\\
$\rm [\alpha /Fe]$ & +0.3 & +0.3 & & +0.3 & +0.3\\
$\rm [C/Fe]$ & --0.3 & --0.3 & & --0.3 & --0.3\\
\enddata
\end{deluxetable}

\begin{deluxetable}{lccccccccc}
\footnotesize
\tablewidth{15truecm}
\tablecaption{Atomic line modeling: excitation potentials (eV), 
oscillator strengths and equivalent widths (m\AA).}
\tablehead{
\colhead{Line}&
\colhead{$\chi $(eV)}& 
\colhead{} &
\colhead{log--gf}& \multicolumn{2}{c}{EW (m\AA)} & 
\colhead{} &
\colhead{log--gf}& \multicolumn{2}{c}{EW (m\AA)} \\ 
\colhead{} &
\colhead{} &
\colhead{} &
\colhead{Kurucz$^a$} &
\colhead{mod1$^b$}&
\colhead{mod2$^c$}&
\colhead{} &
\colhead{MB99} &
\colhead{mod1$^b$}&
\colhead{mod2$^c$} \\
}
\startdata
Ca~$\lambda $1.61508 & 4.53 && -0.190  &304 &284 && -0.34 & 294 & 271 \\
Fe~$\lambda $1.61532 & 5.35 && -0.821  &252 &237 && -0.82 & 252 & 237 \\
Fe~$\lambda $1.55317 & 5.64 && -0.357  &202 &206 && -0.73 & 168 & 169 \\
Mg~$\lambda $1.57658 & 5.93 &&  0.380  &424 &437 &&  0.07 & 391 & 405 \\
Si~$\lambda $1.58884 & 5.08 && -0.030  &504 &485 && -0.25 & 495 & 474 \\
\enddata
\tablecomments{
~~~~~~~~~~~~\\
$^a$ For the Ca line the log--gf value is taken from BG73.\\ 
$^b$ Best fit model to the observed spectra of Liller~1 (cf. Table~2).\\
$^c$ Best fit model to the observed spectra of NGC~6553 (cf. Table~2).\\
}
\end{deluxetable}

\clearpage
\begin{figure}
\epsscale{1.0}
\plotone{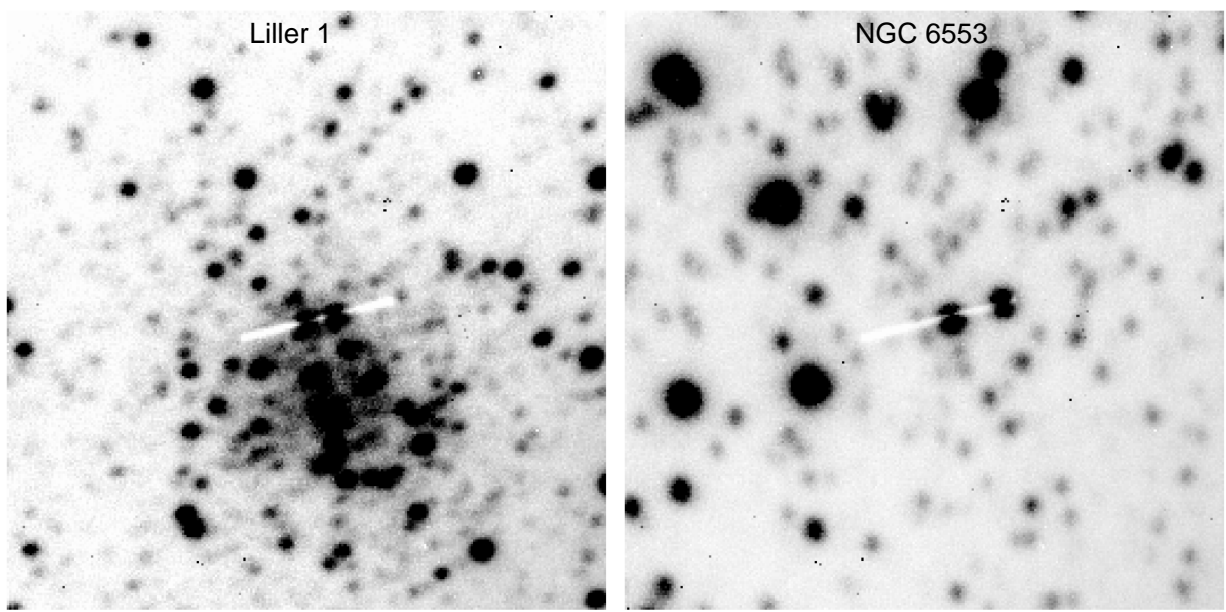}
\caption{
H band images of the fields in the cores of Liller~1 and NGC~6553 as
imaged by the slit viewing camera (SCAM) of NIRSPEC.
The field of view is 46\arcsec on a side (North up,
East to the left), and the image scale is
$0\farcs183$$~pixel^{-1}$; the slit is 12\arcsec\ long.
}
\end{figure}

\clearpage
\begin{figure}
\epsscale{1.0}
\plotone{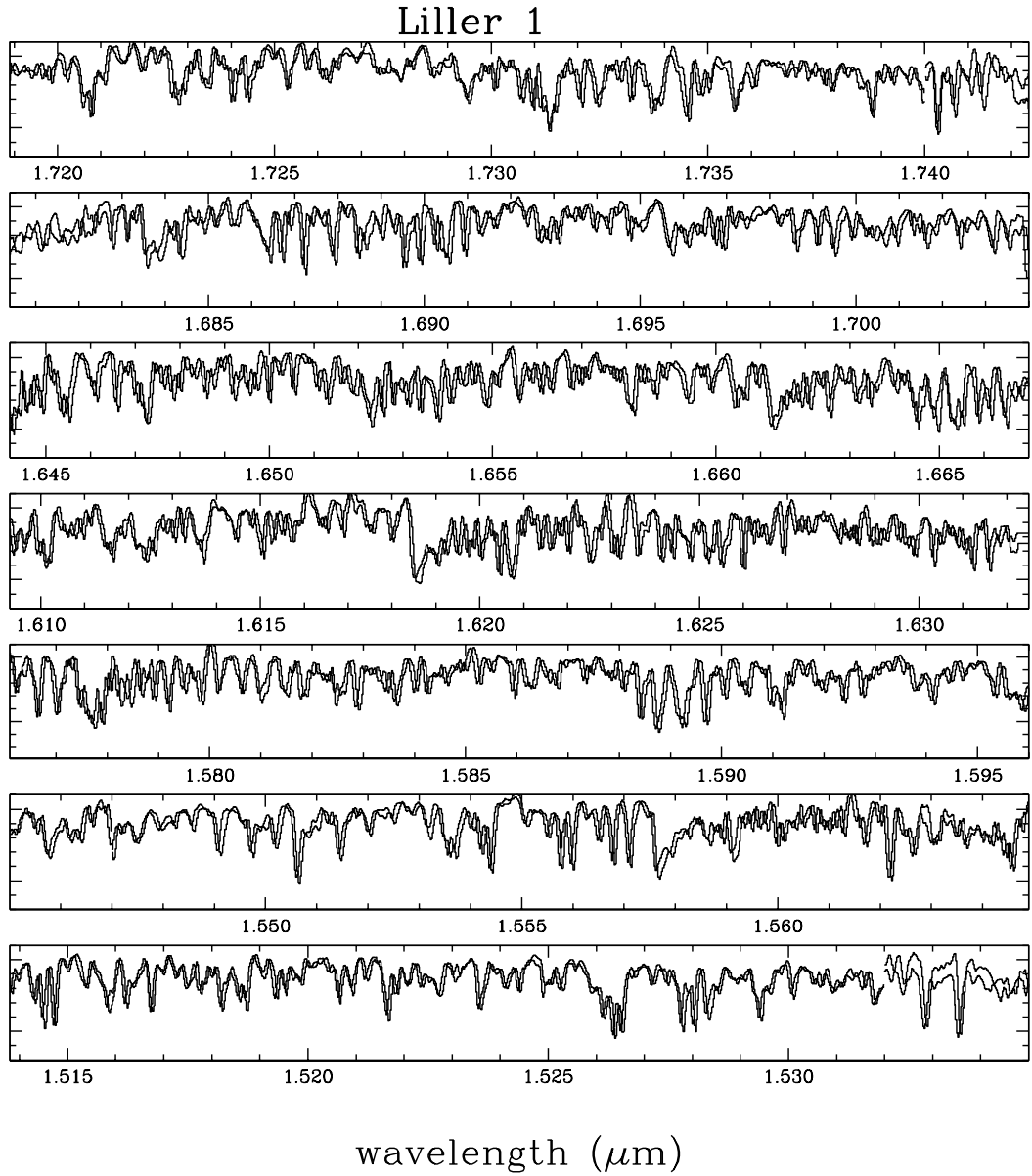}
\caption{
Normalized echelle spectra in the 1.51-1.75 $\mu$m wavelength range
of the two giant stars in Liller~1.}
\end{figure}

\clearpage
\begin{figure}
\epsscale{1.0}
\plotone{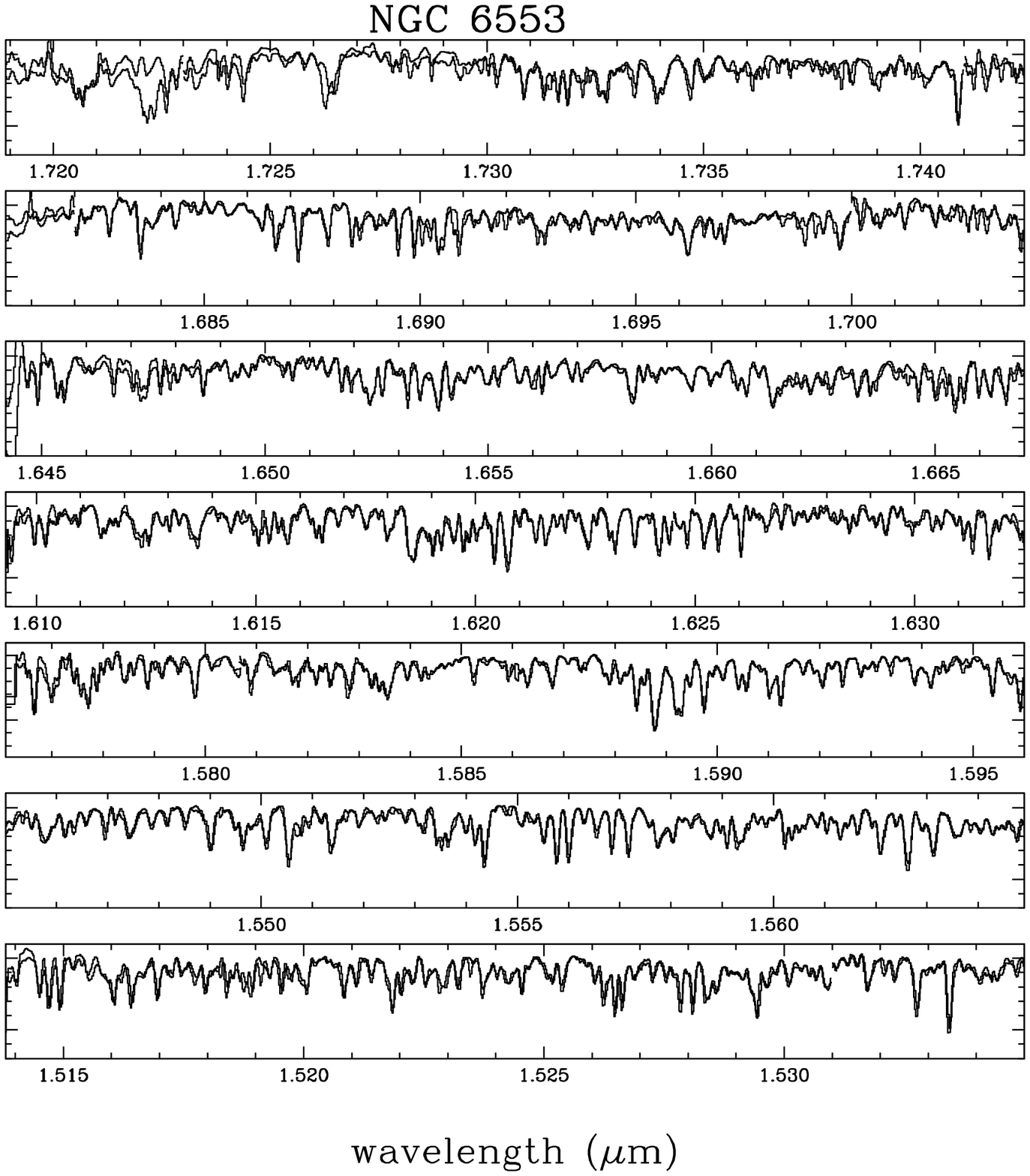}
\caption{
As in Fig.~2 but for the two giant stars in NGC~6553.}
\end{figure}

\clearpage
\begin{figure}
\epsscale{1.0}
\plotone{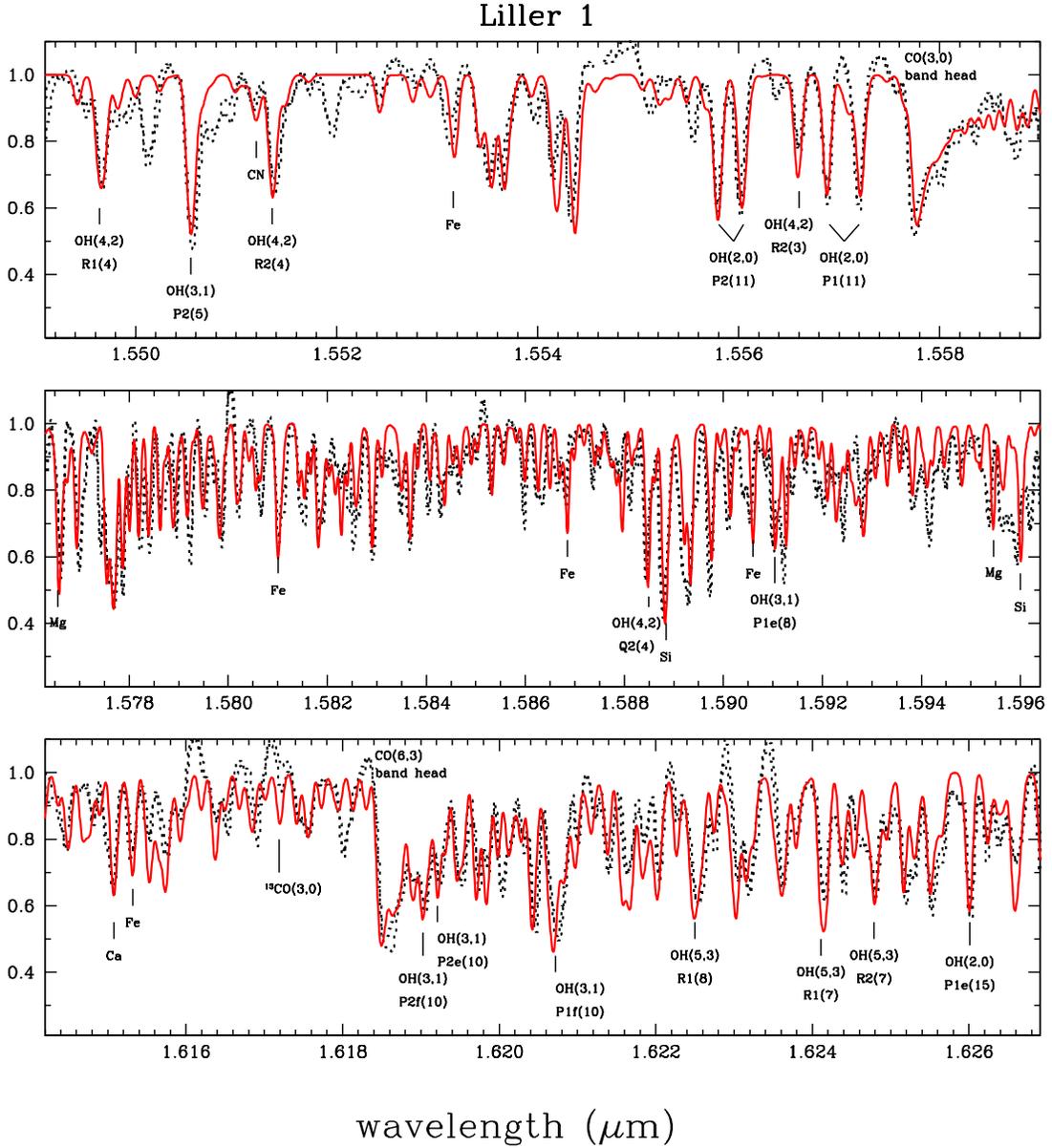}
\caption{
Selected portions of the observed echelle spectra (dotted lines) of the
two giants in Liller~1 with our best fit synthetic spectrum 
(solid line) superimposed. A few important molecular and atomic lines
of interest are marked.
Most of these observed feature are well fitted by our synthetic spectrum 
and the few major discrepancies can be mainly ascribed to poor 
subtraction of overlying OH emission lines in the sky (see also Sect.~3).
The derived abundances are based on a fit of the synthetic spectrum to all 
three orders of the spectrum.
}
\end{figure}

\clearpage
\begin{figure}
\epsscale{1.0}
\plotone{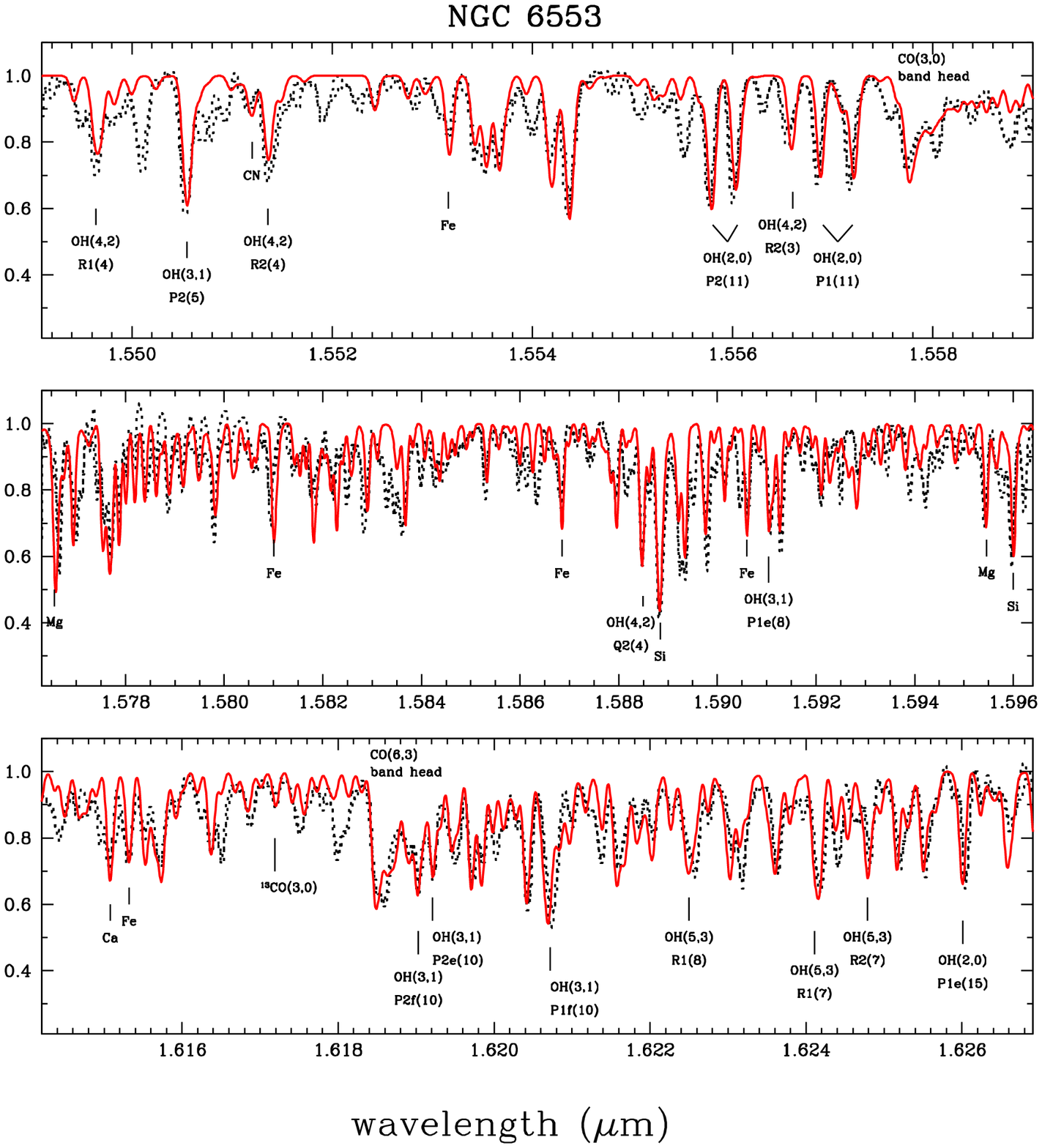}
\caption{
As in Fig.~4 but for the two giant stars in NGC~6553.}
\end{figure}

\clearpage
\begin{figure}
\epsscale{1.0}
\plotone{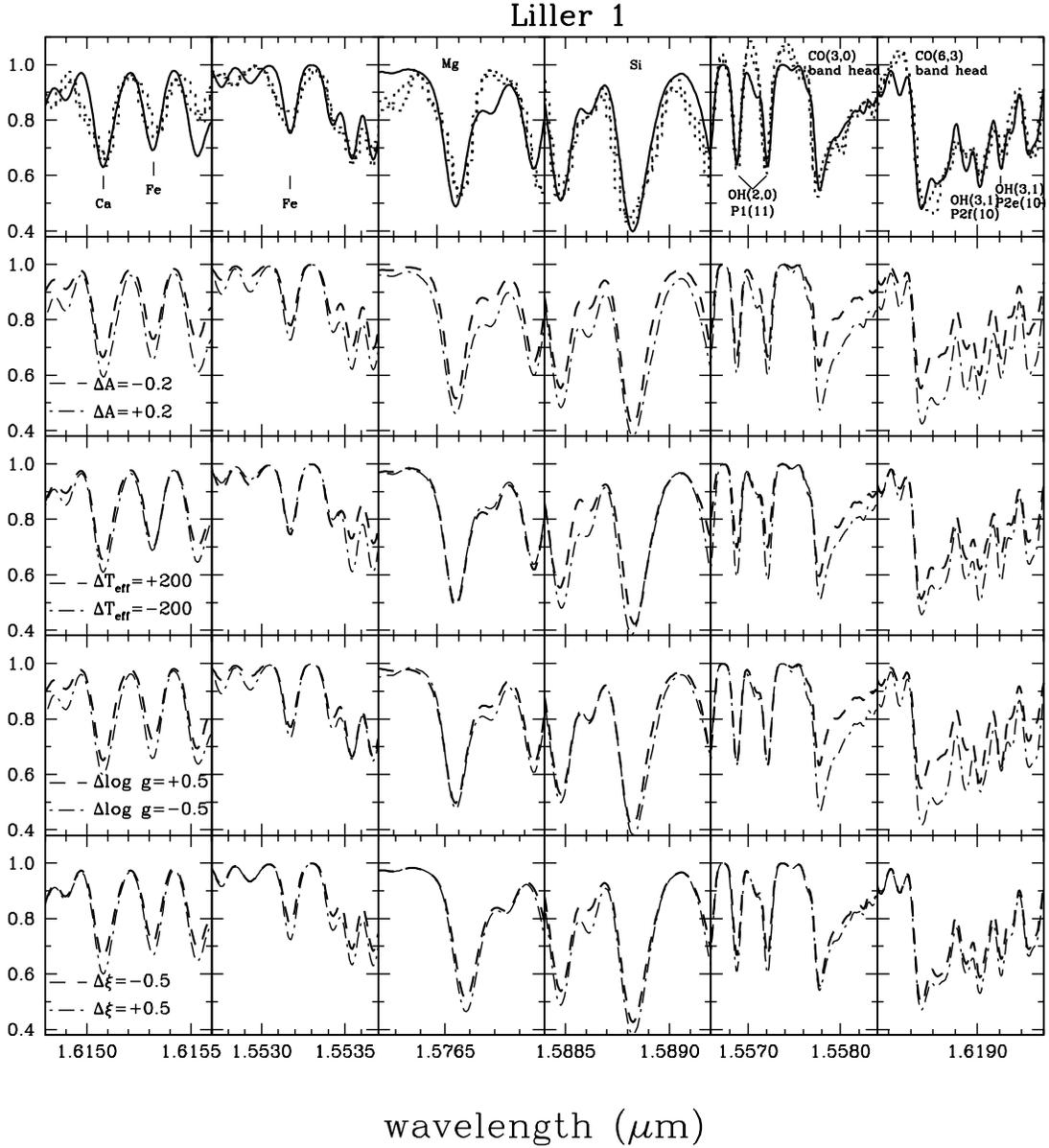}
\caption{Spectra centered on a few atomic and molecular features
of interest.
Top panels: observed spectra of the two giants in Liller~1 (dotted lines)
and our best fit (solid line), using
T$_{\rm eff}$=3800~K, log~g=0.5, $\xi$=2~km~s$^{-1}$, [Fe/H]=--0.3,
[$\alpha$/Fe]=+0.3, [C/Fe]=--0.3 as reference stellar parameters.
Below, from the top to the bottom:
synthetic spectra with $\Delta $A=$\pm$0.2~dex,
$\Delta $T$_{\rm eff}$$\mp$200~K, $\Delta $log~g=$\mp$0.5 and
$\Delta \xi$$\pm$0.5~~km~s$^{-1}$ with respect to the above reference
parameters (cf. also Table~2).
}
\end{figure}

\clearpage
\begin{figure}
\epsscale{1.0}
\plotone{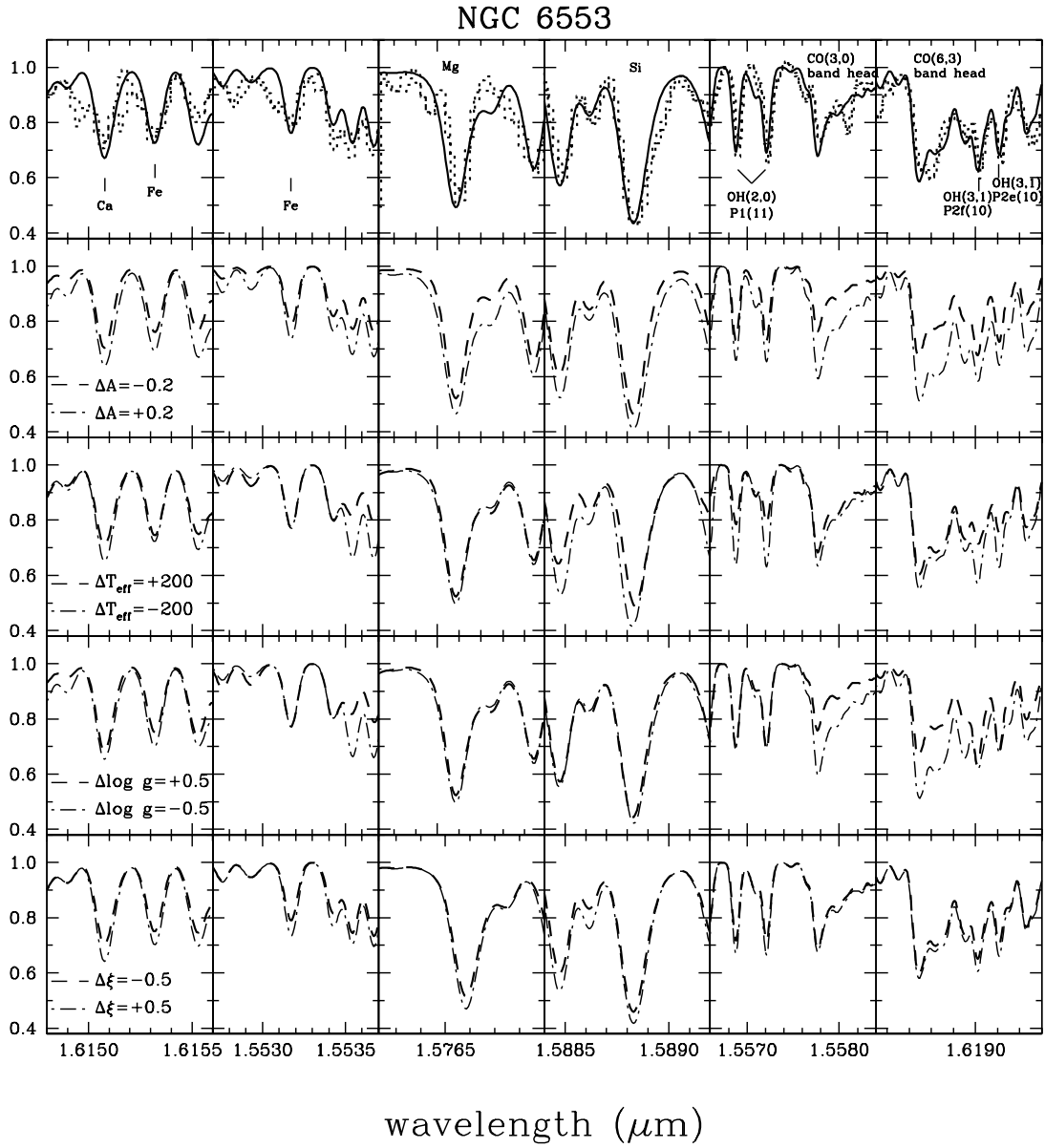}
\caption{ As in Fig.~6, but for the stars in NGC~6553 and
T$_{\rm eff}$=4000~K, log~g=1.0 reference temperature and gravity.}
\end{figure}

\clearpage
\begin{figure}
\epsscale{1.0}
\plotone{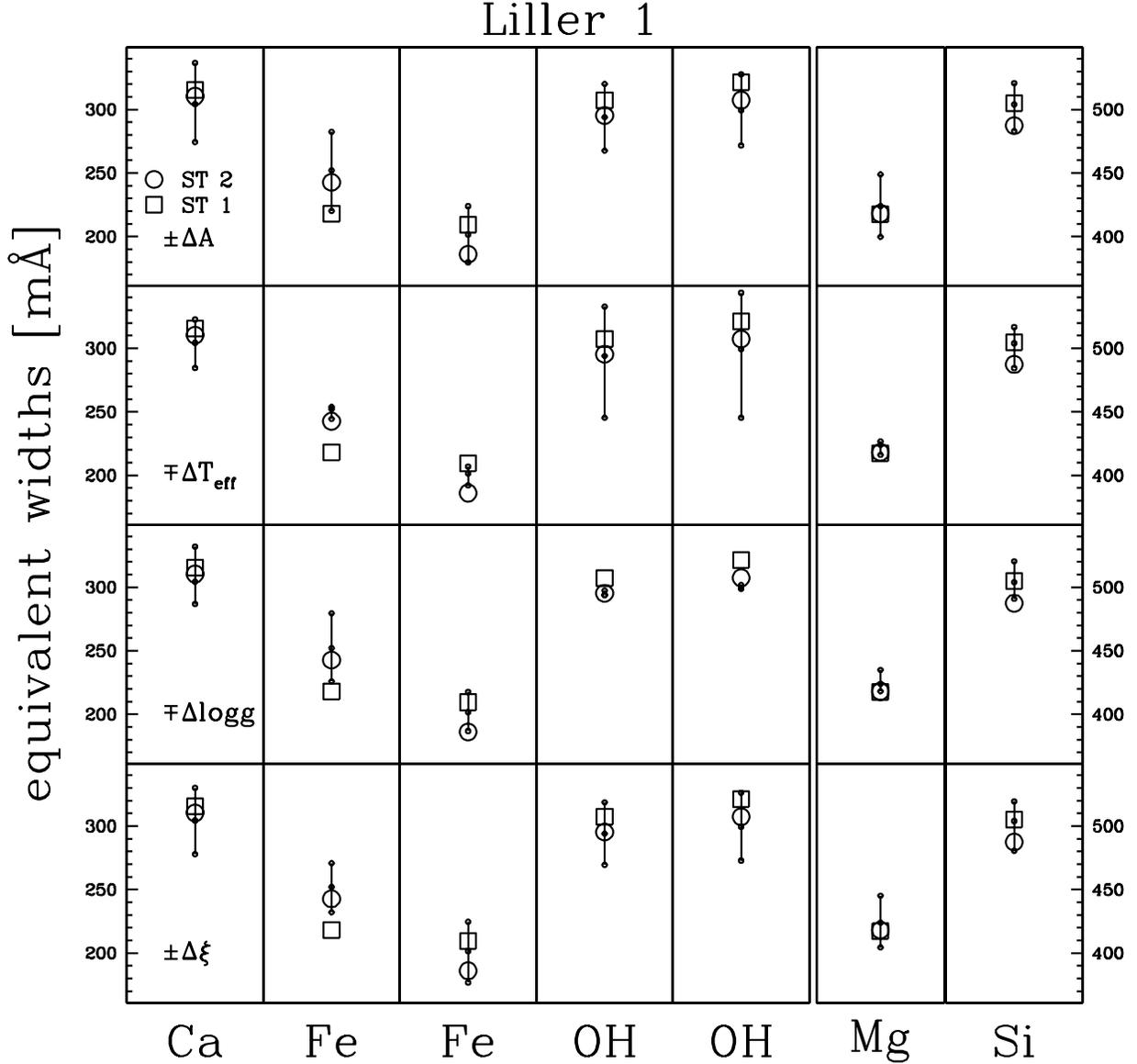}
\caption{Liller~1 : measured equivalent widths for the selected
atomic and molecular
features from the observed (symbols) and synthetic (dots
connected by vertical lines) spectra plotted in Fig.~6
with varying the stellar parameters.
The central dot refers to the models with the
the reference parameters listed in Table~2, while
the top and bottom dots refer to models with
$\Delta $A=$\pm$0.2~dex,
$\Delta $T$_{\rm eff}$$\mp$200~K, $\Delta $log~g=$\mp$0.5 and
$\Delta \xi$$\pm$0.5~~km~s$^{-1}$ with respect to the
reference one. 
}
\end{figure}

\clearpage
\begin{figure}
\epsscale{1.0}
\plotone{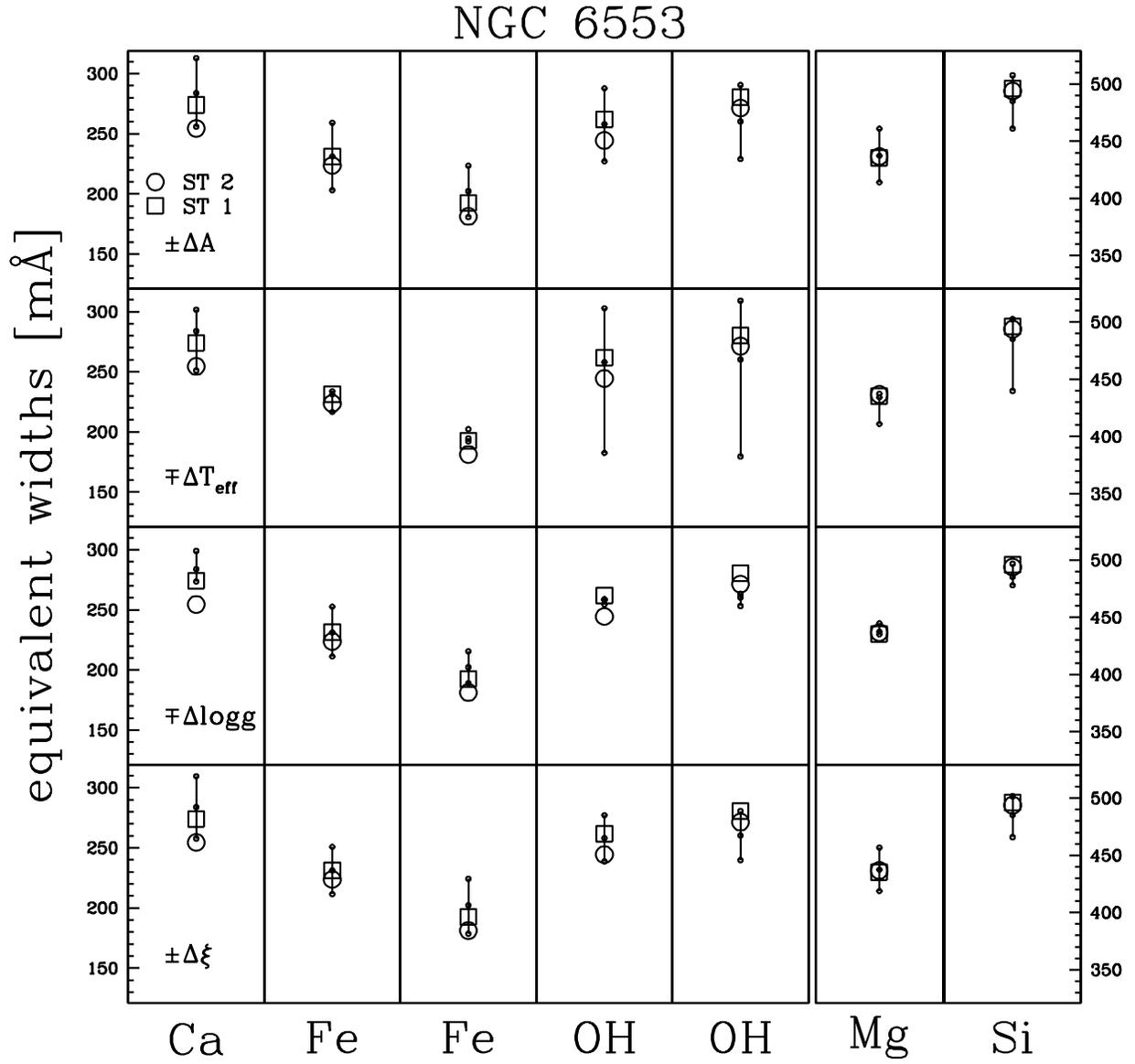}
\caption{ As in Fig.~8, but for the stars and in NGC~6553 
and the synthetic spectra plotted in Fig.~7.
}
\end{figure}

\clearpage
\begin{figure}
\epsscale{1.0}
\plotone{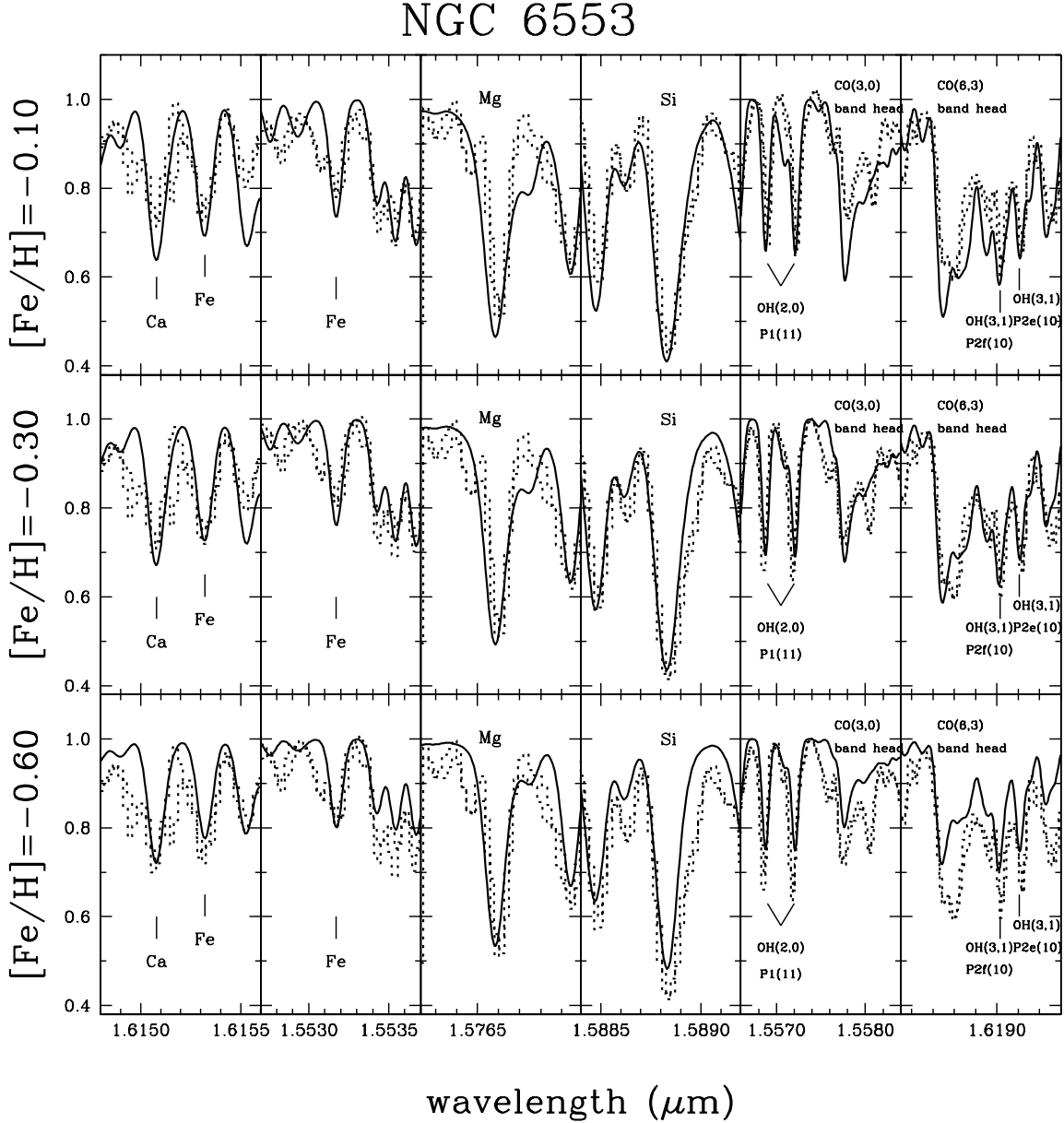}
\caption{ 
Observed spectra of the two giants in NGC~6553 (dotted lines)
and our best fits (full lines) assuming a metal rich composition of
$\rm [Fe/H]=-0.1$ (top panel),
$\rm [Fe/H]=-0.3$ (middle panel) and 
$\rm [Fe/H]=-0.6$ (bottom panel).
The other stellar parameters are as in Table~2.
These plots show that our [Fe/H]=$-0.3$ and [$\alpha$/Fe]=+0.3
abundances (middle panels) are well constrained.  
An abundance significantly lower than ours gives
an especially poor fit to the data.
}
\end{figure}

\clearpage
\begin{figure}
\epsscale{1.0}
\plotone{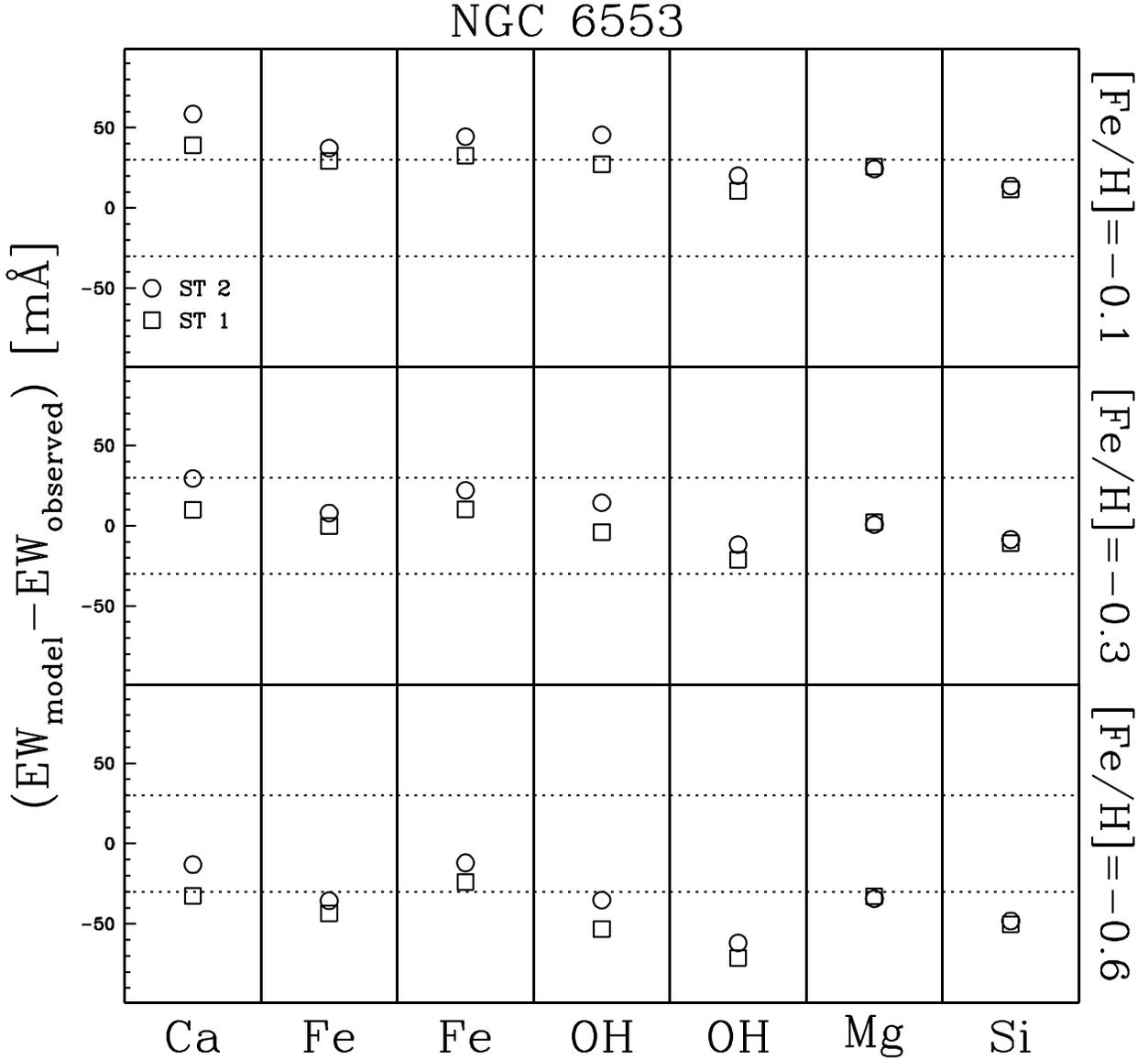}
\caption{ 
Difference between the line equivalent width measured in the models and in 
the observed spectra
of the two giants in NGC~6553, 
assuming a metal rich composition of
$\rm [Fe/H]=-0.1$ (top panel),
$\rm [Fe/H]=-0.3$ (middle panel) and 
$\rm [Fe/H]=-0.6$ (bottom panel).
The other stellar parameters are as in Table~2.
The dotted lines indicate the $\pm 1\sigma$ scatter.
Using our [Fe/H]=$-0.3$ and [$\alpha$/Fe]=+0.3 
abundances (middle panels) 
the difference between the equivalent widths of the model 
and the observed spectra is always $< 1\sigma$, while increases 
expecially 
for an abundance significantly lower than ours.
}
\end{figure}

\end{document}